\documentclass{elsart}
\usepackage{amssymb}
\begin{document}
\begin{frontmatter}

\title{Gauge invariance and weak forces in an isotropic  medium}
\author{A.I. Rez}\footnote{rez@izmiran.rssi.ru},
\author{V.B. Semikoz}\footnote{semikoz@izmiran.rssi.ru}
\address{Institute of Terrestrial Magnetism, Ionosphere and Radio Wave
Propagation of
the Russian Academy of Sciences (IZMIRAN),
142190 Troitsk, Moscow Region, Russia}

\begin{abstract}
We demonstrate that  the neutrino kinetic equation derived
by the standard Bogolyubov method is formally  gauge non-invariant
and give a recipe how to recast it to the gauge invariant form recovering
the standard Lorentz form weak force term which apparently conserves
the lepton current.  The analogy with the QED plasma case
is traced which in asymptotic regions implies substitution of electric
charge $e$ by the induced electric charge $e_{\nu}^{ind}$ of neutrino
within the phase factor connecting
neutrino gauge invariant  and non-invariant distribution functions.

\end{abstract}

\end{frontmatter}

\section{Introduction}
The neutrino Relativistic Kinetic Equation (RKE) is a useful tool to
describe many phenomena in astrophysics and cosmology. In particular,
neutrinos play the most important role for a supernova (SN) burst or in the
lepton asymmetry formation before the primordial nucleosynthesis in the
early universe. The usual motivation to use the RKE approach for
neutrino propagation in a dense matter is stipulated by the account of
neutrino collisions: within a SN neutrinosphere or in the hot
lepton plasma of the early universe before neutrino decoupling.

However, in addition to collision integrals there are self-consistent
weak interaction terms in the neutrino RKE \cite{Semikoz} that are {\it
linear over the Fermi constant $\sim G_F$} and analogous to the Lorentz
force terms for charge particles in the standard Boltzman RKE which in
turn are linear over the electric charge $\sim q$ ($q= - \mid e\mid$ for
electrons).

In the standard kinetics
these self-consistent electromagnetic fields  are well-known to play a
very crucial role.  In collisionless, or Vlasov
approximation, such kinetic equations describe, e.g. thermonuclear
plasmas in laboratory and stars for which an energy exchange between
electromagnetic waves (eigen modes) and charged particles proceeds faster
than via the direct particle collisions with all following issues in
collisionless plasma:  instabilities, heating, etc.

One expects that the self-consistent weak interaction ($\sim G_F$) could
lead for neutrinos to some analogous collective interaction effects, e.g.
to neutrino driven streaming instability in an isotropic plasma
\cite{Bingham}, to generation of magnetic fields
in the early universe or in a SN \cite{Shukla,Brizard}.

Recently neutrino RKE has been rederived along different ways
in \cite{Bingham,Brizard,Bento} and the goals of this letter are: (i) to
establish the conformity of these RKE's with the standard Bogolyubov
approach (Bogolyubov-Born-Green-Kirkwood-Yvon (BBQKY) chains of kinetic
equations) used in \cite{Semikoz}; (ii) to elucidate the physical sense of
the {\it ponderomotive weak force} appeared in effective Lorentz form {\it
meanwhile the neutrino density matrix is gauge invariant with respect to
the corresponding electron current transformation}.

To this end, in section 2 we find that the formal
lepton current nonconservation in the master RKE \cite{Semikoz} stems from
the absense of gauge invariance (see eq.
(\ref{gauge}) below) for the usual in quantum
statistics definition of the gauge
non-invariant distribution $f(\vec{x}_1,\vec{x}_2,t)= Tr
(\hat{\rho}(t)\hat{\Psi}^+(\vec{x}_2)\Psi(\vec{x}_1))$.

Then in section 3 we suggest the recipe of gauge invariance
restoration which allows to derive the Lorentz form of
the weak force term in RKE \cite{Bingham,Bento} and after that in the main
section 4 we find the phase transformation (\ref{phase1}) which connects
the gauge invariant distribution for neutrinos with the
gauge non-invariant one.

The above transformation is fully analogous to the one used in the case
of QED plasma, being based on the same way of inclusion of interaction:
with the self-consistent e/m field $A_{\mu}$ in QED plasma, and with the
self-consistent electron four-current $J_{\mu}$ for neutrinos.
As a result, these transformations turn out to be equivalent with the
accuracy of a change of charges  $e\to e_{\nu}^{ind}$, where
$e^{ind}_{\nu}$ is the induced electric charge of neutrino \cite{OSS}(see
section 4).
This analogy is completed in section 5 where we
recast the weak force term into a form of radiative damping force due to a
plasmon emission by neutrino.

In section 6 we give conclusions and in Appendix we remind some well-known
properties of self-consistent electromagnetic fields in plasma.

\section{Neutrino RKE for gauge non-invariant Wigner distribution}

Neglecting electron spin from the quantum Liouville equation
one finds in the Vlasov approximation the
neutrino Relativistic Kinetic Equation (RKE) as the classical equation for
the gauge non-invariant distribution function $\tilde{f}^{(\nu)}(\vec{q},
\vec{x},t)$ \cite{Semikoz},
\begin{equation}
\label{neutrino3}
\frac{\partial \tilde{f}^{(\nu)}(\vec{q},
\vec{x},t)}{\partial t} + \dot{\vec{x}}\frac{\partial
\tilde{f}^{(\nu)}(\vec{q},\vec{x},t)}{\partial \vec{x}} + \dot{\vec{q}}
\frac{\partial \tilde{f}^{(\nu)}(\vec{q},\vec{x},t)}{\partial \vec{q}}
=0~,
\end{equation}
where $\dot{\vec{x}}=\vec{n}= \vec{q}/q$ is the
velocity of massless neutrino, the derivative $\dot{\vec{q}}$ is given by
\begin{equation}
\label{qforce}
\dot{\vec{q}}= G_F\sqrt{2}c_V\left[\nabla
n^{(e)}(\vec{x},t) - \nabla (\vec{n} \vec{j}^{(e)}(\vec{x},t))\right] ~.
\end{equation}
Here  $ j_{\mu}^{(e)}(\vec{x},t)= (n^{(e)}(\vec{x},t); \vec{
j}^{(e)}(\vec{x},t) )=\int
(\mathrm{d}^3p/(2\pi)^3)(p_{\mu}/\varepsilon_p)f^{(e)}(\vec{p},
\vec{x},t)\\\equiv J^{(e)}_{\mu}(\vec{x},t)/e$ is the four-vector of
the electron current density devided on the electron charge $e = - \mid
e\mid$; $c_V= 2\xi \pm 0.5$ is the weak vector coupling (upper sign for
electron neutrinos), $\xi= \sin^2\theta_W\simeq 0.23$ is the Weinberg
parameter.

Obviously, due to the second term in (\ref{qforce}) the RKE
(\ref{neutrino3}) does not obey the neutrino current conservation law, $$
\frac{\partial j^{(\nu)}_{\mu}(\vec{x},t)}{\partial x_{\mu}}\neq 0~, $$
where $j^{(\nu)}_{\mu}(\vec{x},t)= \int
\mathrm{d}^3q(q_{\mu}/q)\tilde{f}^{(\nu)}(\vec{q},\vec{x},t)/(2\pi)^3$ is
the neutrino four-current {\it in medium}.

First, note that RKE (\ref{neutrino3}) can be derived from the canonical equation
$\partial \tilde{f}^{(\nu)}/\partial t + \{H,\tilde{f}^{(\nu)} \}=0$
with use of the neutrino Hamiltonian {\it in a medium} \cite{Bingham}
$$
H=H_0 + V_{eff}= \sqrt{( \vec{Q}- G_F\sqrt{2}c_V\vec{j}^{(e)}(\vec{x},t))^2  } + G_F\sqrt{2}c_Vn^{(e)}(\vec{x},t)~,
$$
Here the kinematical momentum $\vec{q}$ of massless neutrino is connected with the canonical one, $\vec{Q}$, as
\begin{equation}
\label{conjugate}
\vec{q}= \vec{Q} - G_F\sqrt{2}c_V\vec{j}^{(e)}(\vec{x},t)~,
\end{equation}
and canonical definitions $\vec{n}=\vec{q}/q= \partial H/\partial
\vec{Q}\equiv \partial H/\partial \vec{q}$, $\dot{q}^i= -
\partial H/\partial x_i=\nabla H$ lead to (\ref{neutrino3}).

The origin of the non-conservation of the lepton current
seen from (\ref{neutrino3}) is {\it the absence of the invariance of the Hamiltonian $H$}
(and, hence, of the neutrino RKE) with respect to the gauge transformation
of the electron current
\begin{equation}
\label{gauge}
j^{(e)}_{\mu}(\vec{x},t)\to
j^{(e)}_{\mu}(\vec{x},t) -
\partial_{\mu}\chi  (\vec{x},t)~,
\end{equation}
where an  arbitrary function $\chi (\vec{x},t)$ should also  obey
d'Alambert equation\\ $\partial^{\mu}\partial_{\mu}\chi(\vec{x},t)=0$.

Note that the invariance of the one-particle neutrino motion equation
in a medium under
the same gauge transformation (\ref{gauge}) is equivalent to the neutrino
current conservation too since in the integrand of the action $S= \int
(L_0+L_{int}(\vec{x},t))\mathrm{d}^4x$ there appears an additional (second)
term $$G_Fj_{(\nu)}^{\mu}(\vec{x},t)j_{\mu}^{(e)}(\vec{x},t)\to
G_Fj_{(\nu)}^{\mu}(\vec{x},t)j_{\mu}^{(e)}(\vec{x},t) - G_F
j_{(\nu)}^{\mu}(\vec{x},t)\partial_{\mu} \chi (\vec{x},t)~, $$
which does not contribute to the action, or such
gauge transformation should  not influence the  motion equation
coming from the extremum, $\delta S=0$,
exactly due to $\partial j^{(\nu)}_{\mu}(\vec{x},t)/\partial x_{\mu}=0$.

\section{Neutrino RKE for the gauge invariant Wigner distribution}

The recipe of gauge invariance restoration is the same as in QED plasma
\cite{Fujita}: we should recast (\ref{neutrino3}) for the gauge invariant
Wigner distribution function
\begin{equation}
\label{Fujita1}
f^{(\nu)}(\vec{q},\vec{x},t)=
\tilde{f}^{(\nu)}(\vec{Q},\vec{x},t)~,
\end{equation}
where the latter obeys
the same RKE (\ref{neutrino3}) but with the substitution of the
kinematical momentum $\vec{q}$ by the canonical one (\ref{conjugate}),
\[
\frac{\partial \tilde{f}^{(\nu)}(\vec{Q},\vec{x},t)}{\partial t} +
\vec{n}\frac{ \partial \tilde{f}^{(\nu)}(\vec{Q},\vec{x},t) }{\partial \vec{x}}
+ \dot{\vec{Q}}\frac{\partial
\tilde{f}^{(\nu)}(\vec{Q},\vec{x},t)}{\partial \vec{Q}}=0~.
\]
Accounting for (\ref{qforce}), (\ref{conjugate}), the total time
derivative $\mathrm{d}\vec{j}^{(e)}(\vec{x},t)/\mathrm{d}t= \partial
\vec{j}^{(e)}(\vec{x},t)/\partial t + (\vec{n}\nabla)
\vec{j}^{(e)}(\vec{x},t)$ with the identity
\[
(\vec{n}\nabla)\vec{j}^{(e)}(\vec{x},t) -
\nabla(\vec{n}\vec{j}^{(e)}(\vec{x},t))\equiv -
[\vec{n}\times \nabla\times \vec{j}^{(e)}(\vec{x},t)]~,
\]
and using the recipe
(\ref{Fujita1}) one can easily check that the RKE above takes the form
\cite{Bingham}
\begin{eqnarray}
\label{neutrino1}&&\frac{\partial
f^{(\nu)}(\vec{q},\vec{x}, t )}{\partial t} + \vec{n}\frac{\partial
f^{(\nu)}(\vec{q},\vec{x}, t )}{\partial \vec{x}} +
F_{j\mu}^{(V)}(\vec{x},t)\frac{q^{\mu}}{\varepsilon_q}\frac{\partial
f^{(\nu)}(\vec{q},\vec{x},t)}{\partial q_j}= 0~,
\end{eqnarray}
where the antisymmetric tensor $F_{jk}^{(V)}(\vec{x},t)$ entering the
effective Lorentz force is given by the weak vector current,
\begin{eqnarray} \label{tensors}
&&F_{j0}^{(V)}(\vec{x},t)/G_F\sqrt{2}c_V= -\nabla_j
n^{(e)}(\vec{x},t) - \frac{\partial
j^{(e)}_j(\vec{x},t)}{\partial
t}~,\nonumber\\
&&F_{jk}^{(V)}(\vec{x},t)/G_F\sqrt{2}c_V=
e_{jkl}(\nabla\times
\vec{j}^{(e)}(\vec{x},t))_l~,
\end{eqnarray}
and in accordance with (\ref{conjugate}) we changed the derivative
$\partial /\partial \vec{Q}\to \partial /\partial \vec{q}$.

Obviously, the tensor (\ref{tensors}) (and thus the whole RKE
(\ref{neutrino1})) is invariant with respect to the transformation
(\ref{gauge}) and obeys the continuity equation, or the neutrino current
$j_{\mu}^{(\nu)}(\vec{x},t)= \int
(\mathrm{d}^3q/(2\pi)^3)(q_{\mu}/q)f^{(\nu)}(\vec{p}, \vec{x}, t)$ is
conserved, \begin{equation} \label{neutrinocurrent} \frac{\partial
j_{\mu}^{(\nu)}(\vec{x},t)}{\partial x_{\mu}}=0~.
\end{equation}

So far we did not obtain any new formulae.
However, we argue that the above formal recipe (\ref{Fujita1}) is a simple
consequence of the gauge invariance under the transformation
(\ref{gauge}). In the next section we try to elucidate the physical sense
of such invariance in plasma and its connection with the important
definition of the gauge invariant Wigner distribution in SM
(\ref{Fujita1}).

\section{Gauge invariant Wigner distribution  and induced electric charge of
neutrino}
Let us remind the definitions and the physical sense of the gauge
invariant Wigner distribution functions in QED plasma \cite{Peletminsky},
\begin{eqnarray}
\label{chargeWigner}
f^{(e)}(\vec{p},\vec{x},t) & = & \tilde{f}^{(e)}(\vec{p}+
e\vec{A}(\vec{x},t) , \vec{x},t) \nonumber\\
& = & \int \mathrm{d}^3ye^{i\vec{p}\vec{y}}f^{(e)}(\vec{x}-
\vec{y}/2,\vec{x} + \vec{y}/2,t)~,
\end{eqnarray}
where the gauge invariant
distribution function in the coordinate representation
$f^{(e)}(\vec{x}_1,\vec{x}_2,t)$ is connected with the gauge non-invariant
$\tilde{f}^{(e)}(\vec{x}_1,\vec{x}_2,t)=
Tr\left(\hat{\rho}(t)\hat{\Psi}^{(e)+}(\vec{x}_2)\hat{\Psi}^{(e)}(\vec{x}_1)\right)$
by the important phase factor \cite{Fujita}:
\begin{eqnarray}
\label{phase}
f^{(e)}(\vec{x}_1,\vec{x}_2,t) & = & \exp \left[ie(\vec{x}_2-\vec{x}_1)
\int_0^1\mathrm{d}\xi\vec{A}\Bigl(\vec{x}_2
+ \xi(\vec{x}_1-
\vec{x}_2),t\Bigr)\right]\times \nonumber\\
& \times & \tilde{f}^{(e)}(\vec{
x}_1,\vec{x}_2,t)~.
\end{eqnarray}
Namely due to this phase factor the
distribution (\ref{phase}) is invariant under the standard gauge
transformation (with an arbitrary gauge function $\chi (\vec{x},t)$
obeying the d'Alambert equation),
\begin{eqnarray}
\label{standard}
&&\hat{\Psi}^{(e)}(\vec{x}_1)\to e^{-ie\chi(\vec{x}_1,t)}
\hat{\Psi}^{(e)}(\vec{x}_1)~,\nonumber\\
&& \hat{\Psi}^{(e)+}(\vec{x}_2)\to
e^{+ie\chi(\vec{x}_2,t)} \hat{\Psi}^{(e)}(\vec{x}_2)~,\nonumber\\
&& \vec{ A}\left(\vec{x}_2 + \xi(\vec{x}_1- \vec{x}_2),t\right)\to \vec{
A}\left(\vec{x}_2 + \xi(\vec{x}_1- \vec{x}_2),t\right) \nonumber\\
& - & \frac{\partial \chi \left(\vec{x}_2 + \xi(\vec{x}_1- \vec{
x}_2),t\right)}{\partial \vec{x}_2}~,
\end{eqnarray}
or, equivalently,
this arbitrary phase $\chi (\vec{x},t)$ cancels in (\ref{phase}).  Such
invariance is crucial for macroscopic physics since it provides the
physical sense of the Wigner function (\ref{chargeWigner}) and the
conservation of the macroscopic electric current.

Really, as in the case of
neutrino RKE (\ref{neutrino3}), the kinetic equation for the gauge
non-invariant distribution of charged particles
$\tilde{f}^{(e)}(\vec{x}_1,\vec{x}_2,t)$ derived from the quantum
Liouville equation by the same Bogolyubov method {\it does not obey
electric current conservation}. This is because the force term depends on
the electromagnetic potentials $A_{\mu}(\vec{x},t)$ which do not enter as
combinations expressed via field strengths, $\vec{E}$, $\vec{B}$
\cite{Peletminsky}.  The recasting of such RKE for the gauge-invariant
distribution (\ref{chargeWigner}) allows to obtain the usual Lorentz form
of the force term in the standard Boltzman equation for charged particles
\cite{Peletminsky}:
\[
\frac{\partial f^{(e)}(\vec{p},\vec{x},t)}{\partial t}+
\vec{v}\frac{\partial f^{(e)}(\vec{p},\vec{x},t)}{\partial \vec{x}}+
e\left(\vec{E}(\vec{x},t) + [\vec{v}\times
\vec{B}(\vec{x},t)]\right)\frac{\partial
f^{(e)}(\vec{p},\vec{x},t)}{\partial \vec{p}}=0~,
\]
for which, of course,
the electric current\\ $j^{(e)}(\vec{x},t)= \int
\mathrm{d}^3p(p_{\mu}/\varepsilon_p)f^{(e)}(\vec{p},\vec{x},t)/(2\pi)^3$ is
conserved, $\partial j_{\mu}^{(e)}/\partial x_{\mu}=0$.

Note that through the whole text we use exactly this standard RKE for
electrons neglecting their mutual weak interactions.

Hence, in analogy with the QED plasma definition (\ref{phase})
we should reformulate (\ref{Fujita1}), which leads to the neutrino current
conservation in final RKE (\ref{neutrino1}), for the gauge invariant
distribution in the coordinate representation, $f^{(\nu)}(\vec{x}-
\vec{y}/2,\vec{x}+ \vec{y}/2,t)=\int \mathrm{d}^3qe^{-i\vec{q}\vec{y}}
f^{(\nu)}(\vec{q},\vec{x},t)/(2\pi)^3$ .

To this end, comparing the neutrino canonical momentum (\ref{conjugate})
with the well-known $\vec{p}= \vec{P}- e \vec{A}(\vec{x},t) $ in the case
of QED plasma, we find the important (weak ) phase factor that connects the
gauge invariant distribution $f^{(\nu)}(\vec{x}_1,\vec{x}_2,t)$ with the
gauge non-invariant $\tilde{f}^{(\nu)}(\vec{x}_1,\vec{x}_2,t)$,
\begin{eqnarray}
\label{phase1}
f^{(\nu)}(\vec{x}_1,\vec{x}_2,t) & = & \exp
\left[iG_F\sqrt{2}c_V(\vec{x}_2-\vec{x}_1)
\int_0^1\mathrm{d}\xi\vec{j}^{(e)}\left(\vec{x}_2 + \xi(\vec{x}_1-
\vec{x}_2),t\right)\right] \nonumber\\
                                 & \times &
\tilde{f}^{(\nu)}(\vec{x}_1, \vec{x}_2,t)~.
\end{eqnarray}
Our goal here is
the explanation of the gauge invariance of this function via the gauge
transformation of electron current in medium (\ref{gauge}) with an
appropriate {\it electromagnetic formfactor of neutrino in plasma} instead
of the electron charge $e$ in the standard QED transformation
(\ref{standard}).

Remembering the definition of the gauge non-invariant distribution
$\tilde{f}^{(\nu)}(\vec{x}_1,\vec{x}_2,t) \\=
Tr\left(\hat{\rho}(t)\hat{\Psi}^{(\nu)+}(\vec{x}_2)
\hat{\Psi}^{(\nu)}(\vec{x}_1)\right)$ (compare with the electron case
before (\ref{phase})) we find in the completed form how the gauge
invariance (\ref{gauge}) with an arbitrary gauge $\chi (\vec{x},t)$
is manifested in the neutrino kinetics :
\begin{eqnarray}
\label{gauge1}
&&\hat{\Psi}^{(\nu)}(\vec{x}_1)\to \exp
\left(-iG_F\sqrt{2}c_V\chi(\vec{x}_1,t)\right)
\hat{\Psi}^{(\nu)}(\vec{x}_1)~,\nonumber\\
&&\hat{\Psi}^{(\nu)+}(\vec{x}_2)\to \exp
\left(+iG_F\sqrt{2}c_V\chi(\vec{x}_2,t)\right)
\hat{\Psi}^{(\nu)}(\vec{x}_2)~,\nonumber\\
&& \vec{j}^{(e)}\Bigl(\vec{x}_2
+ \xi(\vec{x}_1- \vec{x}_2),t\Bigr)\to \vec{j}^{(e)}\Bigl(\vec{x}_2 +
\xi(\vec{x}_1- \vec{x}_2),t\Bigr)  \nonumber\\
& - & \frac{\partial \chi
\left(\vec{x}_2 + \xi(\vec{x}_1- \vec{x}_2),t\right)}{\partial
\vec{x}_2}~.
\end{eqnarray}
It is easy to check the cancellation of $\chi$ in the phase factor in
(\ref{phase1}).

On the one hand, this transformation provides
the gauge invariance of the neutrino distribution (\ref{phase1}) and
automatically the invariance of the Wigner function
$f^{(\nu)}(\vec{q},\vec{x},t)$ (\ref{Fujita1}) resulting in the neutrino
current conservation (\ref{neutrinocurrent}).

On the other hand, in an isotropic plasma the electric current\\
$J^{(e)}_i= -\mid e\mid j_i^{(e)}$ is the induced one,
\[
J_i^{(e)}\Bigl(\vec{x}_2 + \xi(\vec{x}_1-\vec{x}_2),t\Bigr)=\int
\frac{\mathrm{d}^4Q}{(2\pi)^4}e^{-i\omega t + i\vec{k}\Bigl(\vec{x}_2 +
\xi(\vec{x}_1- \vec{x}_2)\Bigr)}\Pi_{i\mu}(\omega,\vec{k})A^{\mu}(\omega,
\vec{k})~,
\]
i.e. the phase factor in (\ref{phase1}) takes the form
\begin{eqnarray}
\label{phase2}
&&\exp \Bigl[-\frac{iG_F\sqrt{2}c_V\mid e\mid}{4\pi
\alpha}y_i\times\int_0^1\mathrm{d}\xi\int \frac{\mathrm{d}^4Q}{(2\pi)^4}
e^{-i\omega t + i\vec{k}(\vec{x}_2 - \xi\vec{y}) }\Bigl
(\frac{Q^2}{k^2}(\varepsilon_l -1)k_ik_j  \nonumber\\ & + &
\omega^2(\delta_{ij}- \frac{k_ik_j}{k^2})(\varepsilon_{tr}-1)\Bigr
)A_j(\omega, \vec{k}) \Bigr]~.  \end{eqnarray} Here $Q_{\mu}= (\omega,
\vec{k})$ is the plasmon four-vector; $\varepsilon_{l,tr}(\omega, k)$ are
longitudinal and transversal permittivities in an isotropic plasma;
$\vec{y}= \vec{x}_2 - \vec{x}_1$.

In the two limiting cases: (i) quasistatic electric field ($\omega\ll
k \langle v \rangle \leq k$, $\varepsilon_l-1\approx (kr_D)^{-2}$,
$\omega^2(\varepsilon_{tr}-1)\to 0$) and (ii) high-frequency
electromagnetic field ($\omega\gg k \langle v \rangle $, $Q^2\to 0$,
$\varepsilon_{tr} - 1\approx -\omega_p^2/\omega^2$ ) the phase factor
(\ref{phase2}) takes the
form analogous to the QED plasma result (\ref{phase}):
\begin{eqnarray}
\label{phase3}
\exp \left[-ie_{\nu}^{ind}(\vec{x}_2-\vec{x}_1)
\int_0^1\mathrm{d}\xi\vec{A}\Bigl(\vec{x}_2
+ \xi(\vec{x}_1-
\vec{x}_2),t\Bigr)\right]~,
\end{eqnarray}
where the electromagnetic field splits (see Appendix) into the longitudinal
$\vec{A}\Bigl(\vec{x}_2 + \xi(\vec{x}_1- \vec{x}_2),t\Bigr)=
\vec{A}^{(l)}\Bigl(\vec{x}_2 + \xi(\vec{x}_1- \vec{x}_2),t\Bigr)$ and the
transversal $\vec{A}\Bigl(\vec{x}_2 + \xi(\vec{x}_1- \vec{x}_2),t\Bigr)=
\vec{A}^{(tr)}\Bigl(\vec{x}_2 + \xi(\vec{x}_1- \vec{x}_2),t\Bigr)$
correspondingly multiplied by: (i) either the {\it quasistatic
induced electric charge of neutrino}
or (ii) the high- frequency one \cite{OSS},
\begin{equation}
\label{induced}
e^{ind}_{\nu}= - \frac{\mid e\mid G_Fc_V}{2\pi \alpha
\sqrt{2}r_D^2}~,
\end{equation}
\begin{equation}
\label{induced1}
e^{ind}_{\nu}= - \frac{\mid e\mid
G_Fc_V\omega_p^2}{2\pi \alpha\sqrt{2}}~.
\end{equation}
Here $r_D=\sqrt{T/4\pi \alpha n^{(e)}_0}$ is the Debye radius; $\omega_p=
\sqrt{4\pi \alpha n^{(e)}_0/m_e}$ is the plasma frequency; $T$ and
$n^{(e)}_0$ are the temperature and the mean electron density.

\section{The radiation damping force in isotropic medium}
Here we show that accounting for the lepton current conservation the
final form of neutrino RKE (with an explicit dependence on the
self-consistent electromagnetic fields $\vec{E},~~ \vec{B}$) does not
depend whether we apply initial RKE (\ref{neutrino3}) for the gauge
non-invariant distribution function
$\tilde{f}^{(\nu)}(\vec{q},\vec{x},t)$ \cite{Semikoz}, or the same equation
written in the completed form (\ref{neutrino1}) \cite{Bingham}.

The situation is similar to the case
of standard plasma where the initial RKE for the gauge
non-invariant Wigner distribution
$\tilde{f}^{(e)}(\vec{p},\vec{x},t)$ is often more suitable to
obtain concrete results than Boltzman equation with the Lorentz
force \cite{Peletminsky}. Nevertheless, the electric current should be
written in the gauge invariant form obeying the conservation law,
$\partial j^{(e)}_{\mu}/\partial x_{\mu}=0$ .

Making use in (\ref{neutrino1}) of the standard connection of the
induced electron current $\vec{j}^{(e)}$ with the electric field
$\vec{E}$ (in the Fourier representation), $j_i(\omega,
\vec{k})=(\omega/4\pi
i)\Bigl
[\varepsilon_{ij}(\omega,\vec{k})-\delta_{ij}\Bigr]E_j(\omega,\vec{k})~,$
where the permittivity tensor $\varepsilon_{ij}$ in an isotropic plasma
is given by
\[
\varepsilon_{ij}(\omega,\vec{k})=
\varepsilon_l(\omega,k)\frac{k_ik_j}{k^2}+
\varepsilon_{tr}(\omega,k)\left(\delta_{ij}- \frac{k_ik_j}{k^2}\right)~,
\]
we easily obtain another form of the neutrino RKE
(\ref{neutrino1}),
\begin{eqnarray}
\label{Lorentz}
&& \frac{\partial
f^{(\nu)}(\vec{q}, \vec{x},t)}{\partial t} + \vec{n} \frac{\partial
f^{(\nu)}(\vec{q}, \vec{x},t)}{\partial \vec{x}}
+e\int\frac{\mathrm{d}^4Qe^{-iQx}}{(2\pi)^4} \Bigl
[F_l(\omega,k)\vec{E}_{\parallel}(\omega, \vec{k}) \nonumber\\
& + & F_{tr}(\omega, k))\times \Bigl (\vec{
E}_{\perp}(\omega, \vec{k}) + [\vec{n}\times \vec{B}(\omega, \vec{
k})]\Bigr ) \Bigr ]\frac{\partial f^{(\nu)}(\vec{q}, \vec{x},t)}{\partial
\vec{q}} =0~,
\end{eqnarray}
where $F_l(\omega,k)$ and $F_{tr}(\omega, k)$
are the neutrino electromagnetic formfactors defined in \cite{OSS},
\begin{eqnarray}
\label{formfactors}
&& F_l(\omega,k) =
G_F\sqrt{2}c_V(\varepsilon_l(\omega, k) - 1)Q^2/\alpha~,\nonumber\\
&& F_{tr}(\omega, k)= G_F\sqrt{2}c_V(\varepsilon_{tr}(\omega, k) -
1)\omega^2/\alpha~;
\end{eqnarray}
$\vec{Q}= (\omega, \vec{k})$; $\vec{E}=
\vec{E}_{\parallel}+ \vec{E}_{\perp}$, $\vec{B}$ are the electromagnetic
fields in the dispersive medium, and $\vec{E}_{\parallel}
=\vec{k}(\vec{k}\vec{E})/k^2$.

It is obvious that the third term on the left-hand  side  of  Eq.
(\ref{Lorentz}) is proportional to the force of electromagnetic origin.
Whereas for  a point charge e (when the form factors are equal to unity:
$F_l=F_{tr}= 1$) this term is determined by the Lorentz force, i.e., is
equal to the standard expression $e\Bigl
(\vec{E}(\vec{x},t)+[\vec{n}\vec{B}(\vec{x},t)\Bigr] \partial
f(\vec{q},\vec{x},t)/\partial \vec{q}$, for neutrinos with electromagnetic
structure \cite{OSS}, with allowance for the constant of the weak coupling
to the electric charge, $G_F\sim e^2/M_W^2$, the third term in
(\ref{Lorentz}) is proportional to the radiation damping force ($\sim
e^3$).

The polarization origin of  such  a  force  becomes  clear  after  simple
manipulations in (\ref{Lorentz}) using the explicit expressions
(\ref{formfactors}) for the form factors $F_l$ and $F_{tr}$ in an
isotropic dispersive medium, for which the Fourier integrals can be
completely calculated, and the considered term\cite{Semikoz}
\begin{equation}
\label{polarization1}
\frac{\sqrt{2}G_Fc_V}{e}\left(\frac{\partial^2}{\partial x_j\partial_n} +
n_n\frac{\partial^2}{\partial t\partial x_j} \right )
P_n(\vec{x},t)\frac{\partial f^{\nu}(\vec{q},\vec{x},t)}{\partial q_j}
\end{equation}
is proportional to the second derivative of the polarization vector of the
dispersive medium:
\[
4\pi P_n(\vec{x},t) = D_n(\vec{x},t) - E_n(\vec{x},t)~,
\]
which is equal to the difference between the vectors of the electric
induction, $D_n(\vec{x},t)=\int \mathrm{d}^4x'\varepsilon_{nj}(\vec{x} -
\vec{x}', t - t')E_j(\vec{x'},t)$, and the electric field intensity
$E_n(\vec{x},t)$.

Note that the expression (\ref{polarization1}) for the force is also  valid
for anisotropic media when the permittivity tensor depends, for  example,
on an  external magnetic field. In  vacuum  ($\vec{D}=\vec{E}$),  the
effect corresponding to plasmon emission by a moving neutrino disappears,
i.e., there is no damping  force in (\ref{Lorentz}).

Finally, in special cases of the  excitation  in  a  dispersive  medium  of
electrostatic waves ($\omega\ll k \langle v \rangle $)  or  the
propagation  of  a
high-frequency transverse wave ($\omega\gg k \langle v \rangle $), the term
(\ref{polarization1}) can be represented in the form  of the effective
Lorentz force
\[
e^{ind}_{\nu}\vec{E}_{\parallel}\frac{\partial
f^{(\nu)}(\vec{q},\vec{x},t)}{\partial \vec{q}}~,
\]
\[
e^{ind}_{\nu}\Bigl(\vec{E}_{\perp}+ [\vec{n}\vec{B}]\Bigr)\frac{\partial
f^{(\nu)}(\vec{q},\vec{x},t)}{\partial \vec{q}}~,
\]
which is proportional to the neutrino induced  electric  charges
(\ref{induced}) and (\ref{induced1}), respectively.

\section{Conclusions}
Thus, we proved the neutrino current conservation in a medium as the
consequence of the invariance of the distribution
functions (\ref{phase1}) under the gauge transformation
(\ref{gauge1}) that is similar to the standard one (\ref{standard}) for
the charged particle distribution (\ref{phase}).

Second, the ponderomotive force for neutrinos given by the third term in
RKE (\ref{neutrino1}) is the damping (friction) force
arising due to the neutrino electromagnetic structure (the electromagnetic
vertex $\Gamma_{\mu}(\omega,\vec{k})\sim G_F\Pi_{\mu\nu}
(\omega,\vec{k})\gamma^{\nu}$ \cite{OSS}) and is stipulated by the plasmon
($\check{C}$erenkov) emission in medium, $\nu\to \nu+\gamma^*$,
forbidden in vacuum \cite{Semikoz}.

Third,  the appearance of the same neutrino electromagnetic (loop)
structure as the factor ahead the self-consistent electromagnetic fields
in the phase factor (\ref{phase2}) looks rather natural. In particular
cases of quasistatic (longitudinal) or high-frequency (transversal) fields
there is a complete analogy with the QED plasma case accomplished after
substitution of the electric charge $e$ by $e^{ind}_{\nu}$.

\ack
This work was partly supported by the RFBR grant No. 00-02-16271 .

\appendix{Appendix}

In an isotropic plasma the second-quantized electromagnetic fields are
additive, $\hat{A}_{\mu}= \hat{A}^{(l)}_{\mu}+ A^{(tr)}_{\mu}$, where both
the longitudinal field \[ \hat{A}^{(l)}_{\mu}(\vec{x},t)=\int
\frac{\mathrm{d}^3k}{(2\pi)^3N_l(k)}\left(\varepsilon_{\mu}^{(l)}(k)\hat{a}^{(l)}(k)e^{-iqx}+
\varepsilon_{\mu}^{*(l)}(k)\hat{a}^{(l)+}(k)e^{iqx}\right), \]
and the transversal one
\[
A^{(tr)}_{\mu}(\vec{x},t)=\sum_{\lambda}\int
\frac{\mathrm{d}^3k}{(2\pi)^3N_{tr}(k)}
\left(\varepsilon_{\mu}^{\lambda}(k)\hat{a}^{(tr)}_{\lambda}(k)e^{-iqx}+
\varepsilon_{\mu}^{*\lambda}(k)\hat{a}^{(tr)+}_{\lambda}(k)e^{iqx}\right),
\]
obey the Lorentz gauge, $\partial A_{\mu}(\vec{x},t)/\partial x_{\mu}=0$.
Here the unit polarization vectors,
$\varepsilon_{\mu}\varepsilon^{\mu}= -1$, are given by
$\varepsilon_{\mu}^{(l)}(k)=(k, \omega\hat{k})/\sqrt{Q^2},~~~
\varepsilon_{\mu}^{(\lambda)}(k)=(0,
\vec{\varepsilon}^{(\lambda)}(k))~$, and for the
plasmon four-vector $Q_{\mu}= (\omega, \vec{k})$ obey the standard
conditions for both components
$Q^{\mu}\varepsilon_{\mu}^{(l,\lambda)}(k)=0$, or
$k_i\varepsilon_i^{(\lambda)}=0$ for the transversal part only;
 the normalization factors
$N_{l,tr}(k)$, $N_{l}(k)=\sqrt{Q^2\partial
\rm{Re}~\varepsilon_l(\omega,k)/\partial \omega }$,\\
$N_{tr}(k)=\sqrt{2\omega[\varepsilon_{tr}(\omega,k)+ \omega \partial
\rm{Re}~ \varepsilon_{tr}(\omega,k)/\partial \omega]}$ are given by the
longitudinal ($\varepsilon_l(\omega,k)$) and the transversal
($\varepsilon_{tr}(\omega, k)$) permittivities correspondingly.

Note that in the Fourier representation the 3-vector parts of these
four-potentials $\vec{
A}^{(l,tr)}(\omega,k)=\vec{\varepsilon}^{(l,tr)}a(k)/N^{(l,tr)}$ obey
important {\it opposite} conditions $\vec{
A}^{(l)}(\omega,k)\parallel \vec{k}$ and
$\vec{k}\vec{A}^{(tr)}(\omega,k)=0$ that were used above to recast Eq.
(\ref{phase2}) into Eq.  (\ref{phase3}) which is proportional to  induced
electric charges of neutrino (\ref{induced}), (\ref{induced1})
correspondingly.

\end{document}